\documentclass[12pt]{iopart}
\usepackage{iopams}
\usepackage{graphicx,cite}
\begin{document}

\title[FAST TRACK COMMUNICATION]{Efficiency at maximum power of Feynman's ratchet as a heat engine}

\author{Z C Tu}

\address{Department of Physics, Beijing Normal University, Beijing 100875, China}
\ead{tuzc@bnu.edu.cn}
\begin{abstract}
The maximum power of Feynman's ratchet as a heat engine and the
corresponding efficiency ($\eta_\ast$) are investigated by
optimizing both the internal parameter and the external load. When a
perfect ratchet device (no heat exchange between the ratchet and the
paw via kinetic energy) works between two thermal baths at
temperatures $T_1> T_2$, its efficiency at maximum power is found to
be $\eta_\ast =\eta_C^2 /[\eta_C-(1-\eta_C)\ln(1-\eta_C)]$, where
$\eta_C\equiv 1-T_2/T_1$. This efficiency is slightly higher than
the value $1-\sqrt{T_2/T_1}$ obtained by Curzon and Ahlborn
[\textit{Am. J. Phys.} \textbf{43} (1975) 22] for macroscopic heat
engines. It is also slightly larger than the result $\eta_{SS}\equiv
2\eta_C/(4-\eta_C)$ obtained by Schmiedl and Seifert [\textit{EPL}
\textbf{81} (2008) 20003] for stochastic heat engines working at
small temperature difference, while the evident deviation between
$\eta_\ast$ and $\eta_{SS}$ appears at large temperature difference.
For an imperfect ratchet device in which the heat exchange between
the ratchet and the paw via kinetic energy is non-vanishing, the
efficiency at maximum power decreases with increasing the heat
conductivity.

\end{abstract}

\pacs{05.70.Ln}

{J. Phys. A: Math. Theor. 41, 312003 (2008)}
\maketitle

As is well known, the Carnot efficiency $\eta_C\equiv 1-T_2/T_1$
gives the upper bound for heat engines working between two thermal
baths at temperatures $T_1 > T_2$. However, the engines at the
Carnot efficiency cannot produce output power because the Carnot
cycle requires an infinitely slow process. The cycle should be
speeded up to obtain a finite power. Curzon and Ahlborn \cite{CAAJP}
derived the efficiency, $\eta_{CA}\equiv 1-\sqrt{T_2/T_1}$, at
maximum power for macroscopic heat engines within the framework of
finite-time thermodynamics. The same expression was also obtained
from linear irreversible thermodynamics for perfectly coupled
systems \cite{vdbrk,dcisbj,SanchoPRE06}. It is pointed out that the efficiency at maximum
power might be different from $\eta_{CA}$ for imperfectly coupled
systems \cite{vdbrk,dcisbj,SanchoPRE06}. In a recent paper
\cite{Schmiedl08}, Schmiedl and Seifert investigated cyclic Brownian
heat engines working in a time-dependent harmonic potential and
obtained the efficiency at maximum power, $\eta_{SS}\equiv
2\eta_C/(4-\eta_C)$, within the framework of stochastic
thermodynamics \cite{Sekimoto97,Seifert02,Schmiedl07}.

To illustrate the second law of thermodynamics, Feynman introduced
an imaginary microscopic ratchet device in his famous lectures
\cite{Feynmanl}. Following Feynman's spirit, many models were put
forward, such as on-off ratchets \cite{Prost97}, fluctuating
potential ratchets \cite{Astumian,Magnasco}, temperature ratchets
\cite{Reimann96,BaoJD}, chiral ratchets
\cite{Tu2004,Tu2005,vdbrk08}, and so on, which have potential
applications in biological motors. Thus it is significant to
investigate the efficiency and power of these ratchets. Feynman's
ratchet, as a parental model, attracts researchers' interests
\cite{Jarzynski99,Velasco01,AiBQ,Zhangy} in the nature of things. In
particular, the efficiency at maximum power of Feynman's ratchet was
obtained by optimizing the external load for given internal
parameter in Refs.\cite{Velasco01}-\cite{Zhangy}. However, there is
still lack of a result by optimizing both the internal parameter and
the external load of the ratchet device. In the present paper, we
will analytically derive this result.

\begin{figure}[htp!]
\begin{center}\includegraphics[width=9.5cm]{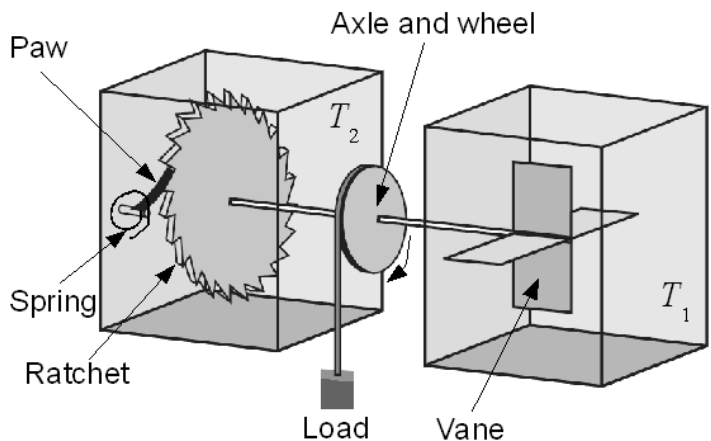}\caption{Feynman's ratchet device.}\label{fig1}
\end{center}\end{figure}

Let us consider Feynman's ratchet device as shown in
figure~\ref{fig1}. It consists of a ratchet, a paw and spring,
vanes, two thermal baths at temperatures $T_1 > T_2$, an axle and
wheel, and a load. For simplicity, we assume that the axle and wheel
is a rigid and frictionless thermal insulator.

Now we follow Feynman's discussion \cite{Feynmanl}. In one-step
forward motion, we must borrow an energy $\epsilon$ to overcome the
elastic energy of spring and then lift the pawl. Assume that the wheel
rotates an angle $\theta$ per step and the load induces a torque $Z$, thus we
also need an additional energy $Z\theta$. Then the total energy that we have
to borrow is $\epsilon+Z\theta$. In fact, we can borrow it from the
hot thermal bath in the form of heat. The rate to get this energy is
\begin{equation}R_F=r_0 e^{-(\epsilon+Z\theta)/T_1},\end{equation}
where $r_0$ is a constant with dimension s$^{-1}$, and the
Boltzmann factor is taken to 1. In this process, the ratchet absorbs
heat $\epsilon+Z\theta$ from the hot thermal bath. A part of this
heat is transduced into work $Z\theta$, and the other is transferred
as heat $\epsilon$ to the cold thermal bath through the interaction
between the ratchet and the paw.

Now let us consider one-step backward motion. To make the wheel backwards, we have to accumulate the energy $\epsilon$ to lift the
pawl high enough so that the ratchet can slip. Here the rate to get
this energy is
\begin{equation}R_B=r_0 e^{-\epsilon/T_2}.\end{equation}
In the backward process, the load does work $Z\theta$. This energy
and the accumulated energy $\epsilon$ are returned to the hot
thermal bath in the form of heat.

In an infinitesimal time interval $\Delta t$, the net work done on
the load by the system may be expressed as
\begin{equation}W =(R_F -R_B)Z\theta\Delta
t.\end{equation} The net heat absorbed from the hot thermal bath via
the potential energy \cite{Parrondo,Astumian99} may be expressed as:
\begin{equation}Q_1^{pot} =(R_F -R_B)(\epsilon+Z\theta)\Delta
t.\label{heat1}\end{equation} Since the ratchet contacts simultaneously with two thermal baths at different temperatures, there may
exist a heat conduction from the hot thermal bath to the cold one
via the kinetic energy \cite{Parrondo,Astumian99}. In time interval
$\Delta t$, it can be expressed as
\begin{equation}Q_1^{kin} =\sigma (T_1-T_2)\Delta
t,\label{heatkin}\end{equation} where $\sigma$ is the heat
conductivity due to the heat exchange between the ratchet and the paw
via kinetic energy. The analysis by Parrondo and Espa\~{n}ol
suggests that $\sigma$ is inversely proportional to the masses
of ratchet and paw \cite{Parrondo}. We first consider an perfect
ratchet device in which the masses of ratchet and paw are infinitely large relative to
the gas molecules full in both thermal baths. In this case,
$\sigma$ and $Q_1^{kin}$ are vanishing. Thus the efficiency can be
defined as
\begin{equation}\eta=W/Q_1^{pot} =Z\theta/(\epsilon+Z\theta).\label{efficd}\end{equation}
The power is defined as \begin{equation}P=W/\Delta t =r_0
Z\theta[e^{-(\epsilon+Z\theta)/T_1}-e^{-\epsilon/T_2}].\label{power1}\end{equation}
We find that $P$ depends on the internal parameter $\epsilon$ and
the external load $Z$. It is easy to tune the external load $Z$. In
fact, $\epsilon$ can also be adjusted by changing the strength of
the spring. We can optimize both $\epsilon$ and $Z$ to achieve the
maximum power. Before doing that, we introduce two dimensionless parameters
$\varepsilon=\epsilon/T_2$ and $z=Z\theta/T_1$. Equations
(\ref{efficd}) and (\ref{power1}) are then respectively transformed into
\begin{equation}\eta=\frac{z}{\varepsilon (1-\eta_C)+z},\label{efficdm}\end{equation} and
\begin{equation}P=r_0 T_1 z[e^{-\varepsilon (1-\eta_C)-z}-e^{-\varepsilon}].\label{powerdm}\end{equation}
Maximizing $P$ with respect to $\varepsilon$ and $z$, we have
\begin{equation}\left\{\begin{array}{l}(1-\eta_C)e^{\varepsilon\eta_C
-z}=1,\\ (1-z)e^{\varepsilon\eta_C -z}=1.
\end{array}
\right.\end{equation} The solution to the above equation is
\begin{equation}\left\{\begin{array}{l}z_\ast=\eta_C,\\ \varepsilon_\ast=1-\eta_C^{-1}\ln(1-\eta_C).
\end{array}
\right.\label{optpare}\end{equation}

Substituting equation (\ref{optpare}) into equation (\ref{efficdm}),
we derive the efficiency at maximum power
\begin{equation}\eta_\ast =\frac{\eta_C^2
}{\eta_C-(1-\eta_C)\ln(1-\eta_C)}.\label{effmaxp}\end{equation} This
is the main result in the present paper. It is not hard to prove
$\eta_{CA}<\eta_\ast<\eta_C$ through simple analysis. Similarly, substituting equation
(\ref{optpare}) into equation (\ref{powerdm}), we obtain the maximum
power \begin{equation}P_{\ast}=\frac{r_0 T_1
\eta_C^2}{e(1-\eta_C)^{1-\eta_C^{-1}}}.\label{maxpower1}
\end{equation}

\begin{figure}[htp!]
\begin{center}\includegraphics[width=9.5cm]{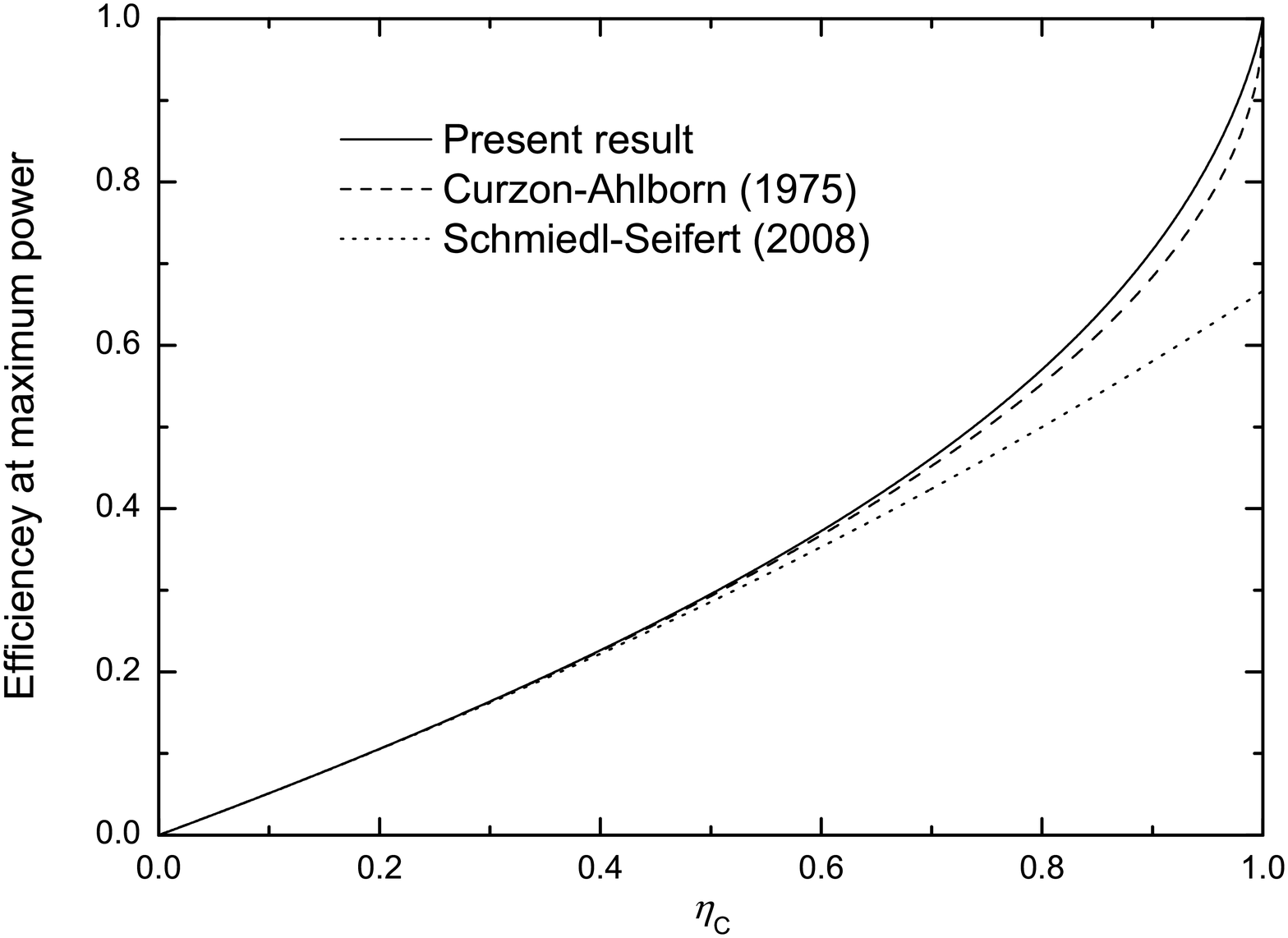}\caption{Comparison between typical results on the efficiency at maximum power obtained by
different research groups.}\label{fig2}\end{center}
\end{figure}

Now we compare the present result with two typical values
$\eta_{CA}$ and $\eta_{SS}$ at various temperatures.
Figure~\ref{fig2} shows the relation between the efficiency at
maximum power and the relative temperature difference,
$(T_1-T_2)/T_1=\eta_C$. We find that the present result (solid line)
is slightly higher than the value $\eta_{CA}\equiv
1-\sqrt{1-\eta_C}$ obtained by Curzon and Ahlborn (dash line) for
macroscopic heat engines, and than the result $\eta_{SS}\equiv 2\eta_C/(4-\eta_C)$
obtained by Schmiedl and Seifert (dot line) for stochastic heat
engines at small relative temperature difference (for example, $\eta_C <0.5$). The evident deviation
between the present result and $\eta_{SS}$ appears merely for $\eta_C >0.5$.

Now we investigate the asymptotic behaviors of $\eta_\ast$,
$\eta_{CA}$, and $\eta_{SS}$ at small relative temperature
difference. Expanding them up to the third order term of $\eta_C$,
we have
\begin{equation}\eta_\ast=\frac{\eta_c}{2} + \frac{\eta_c^2}{8}+ \frac{7\eta_c^3}{96} + O(\eta_c^4),\end{equation}
\begin{equation}\eta_{CA}=\frac{\eta_c}{2} + \frac{\eta_c^2}{8}+ \frac{6\eta_c^3}{96} + O(\eta_c^4),\end{equation}
and
\begin{equation}\eta_{SS}=\frac{\eta_c}{2} + \frac{\eta_c^2}{8}+
\frac{3\eta_c^3}{96} + O(\eta_c^4).\end{equation} These results
deviate from each other at the third order term of relative
temperature difference, which suggests that a universal efficiency
at maximum power, ${\eta_c}/{2} + {\eta_c^2}/{8}$, should exist at
small relative temperature difference.

Up to now, we have only discussed the ideal case that the masses of
ratchet and paw are infinitely large relative to the gas
molecules full in both thermal baths such that $\sigma$ and
$Q_1^{kin}$ are vanishing. In practice, the masses are finite and the ratchet device is imperfect, so
$\sigma$ and $Q_1^{kin}$ are non-vanishing. In this case, the net
heat absorbed from the hot thermal bath is
\cite{Velasco01,Parrondo,Astumian99}:
\begin{equation}Q_1=Q_1^{pot}+Q_1^{kin} =[(R_F -R_B)(\epsilon+Z\theta)+\sigma (T_1-T_2)]\Delta
t.\label{heat11}\end{equation} Correspondingly, the efficiency is
modified as
\begin{equation}\eta^{\prime}=W/Q_1=\frac{z}{\varepsilon (1-\eta_C)+z+e^{-1}\lambda \eta_c [e^{-\varepsilon (1-\eta_C)-z}-e^{-\varepsilon}]^{-1}},\label{efficdmrv}\end{equation}
where $\lambda\equiv e\sigma/r_0$ is the reduced heat conductivity. The power $P$ is independent of the
heat conductivity $\sigma$, so equations
(\ref{powerdm})-(\ref{optpare}) and (\ref{maxpower1}) are unchanged.

\begin{figure}[htp!]
\begin{center}\includegraphics[width=9.5cm]{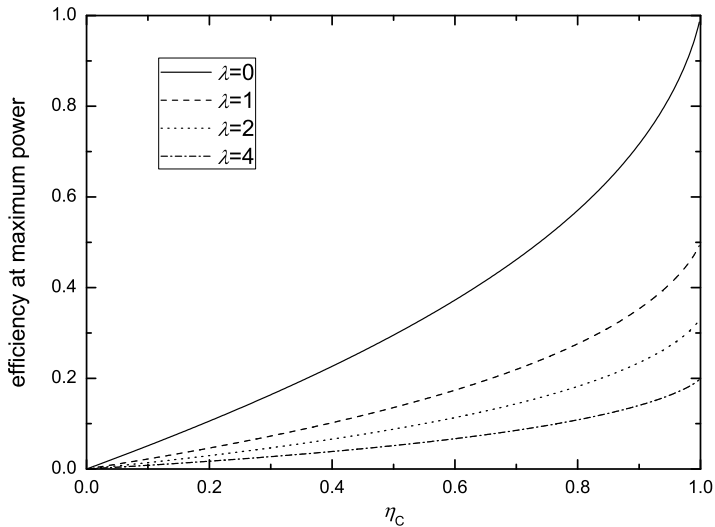}\caption{Efficiency at maximum power for different heat conductivities }\label{fig3}\end{center}
\end{figure}

Substituting equation (\ref{optpare}) into (\ref{efficdmrv}), we
obtain the efficiency at maximum power
\begin{equation}\eta_\ast^{\prime} =\frac{\eta_C^2
}{\eta_C-(1-\eta_C)\ln(1-\eta_C)+\lambda
\eta_C(1-\eta_C)^{1-\eta_C^{-1}}}.\label{effmaxprv}\end{equation} It
is obvious that $\eta_\ast^{\prime}<\eta_\ast$ because $\lambda
\eta_C(1-\eta_C)^{1-\eta_C^{-1}}>0$. The efficiency at maximum power depends on the reduced heat
conductivity $\lambda$. From figure~\ref{fig3}, we see that it
decreases with increasing the reduced heat conductivity.

In summary, we have investigated the efficiency at maximal power of
Feynman's ratchet as a heat engine by optimizing both the internal
parameter and the external load. We have analytically derived the
efficiency at maximum power, equation (\ref{effmaxp}), for perfect
ratchet device, which is slightly higher than the value obtained by
Curzon and Ahlborn. The efficiency at maximum power for the
imperfect ratchet device, equation (\ref{effmaxprv}), decreases with
increasing the reduced heat conductivity. The present result for
perfect ratchet device deviates from the value obtained by Schmiedl
and Seifert merely at large relative temperature difference.

Finally, we have to list a few open problems which should be
addressed in the future work. (i) What is the underlying reason for
the deviation between $\eta_\ast$ and $\eta_{SS}$ at large
temperature difference? A possible conjecture is that there might
not exist a universal formula at large relative temperature
difference. (ii) Zhang \textit{et al.} held a different viewpoint
on $\sigma$ \cite{Zhangy}. They argued that $\sigma= (R_F+R_B)/2$,
which is independent of the masses of ratchet and paw. By adopting
their argument, one can derive the efficiency at maximum power to be
$2\eta_C/[4-\eta_C -2(\eta_C^{-1}-1)\ln(1-\eta_C)]$ which is much
smaller than $\eta_{CA}$, $\eta_{SS}$, and $\eta_{\ast}$ for
$0<\eta_C<1$. Whether $\sigma$ does indeed depend on the rates $R_F$
and $R_B$? (iii) Allahverdyan \textit{et al.} investigated a class
of quantum heat engines consisting of two subsystems interacting
with a work-source and coupled to two separate baths at different
temperatures \cite{Allahverdyan07}. They also found that the
efficiency at maximum power of these quantum heat engines was
slightly larger than $\eta_{CA}$. Is there any relation between
their result and ours?

\textit{Acknowledgments}. The author is grateful for the useful
discussions with Prof. Z. C. Ou-Yang (Chinese Academy of Sciences),
F. Liu (Tsinghua University) and M. Li (Chinese Academy of
Sciences), and for the support from Nature Science Foundation of
China (Grant No. 10704009). The author thanks Dr. K. Fang
(University of California, Los Angeles) and W. H. Zhou (Wuhan
University) for their carefully proofreading the present paper.


\section*{References}
\begin{thebibliography}{10}
\bibitem{CAAJP}Curzon F L and Ahlborn B 1975 \textit{Am. J. Phys.} \textbf{43} 22
\bibitem{vdbrk}van den Broeck C 2005 \textit{Phys. Rev. Lett.} \textbf{95} 190602
\bibitem{dcisbj}de Cisneros B J and Hern\'{a}ndez A C 2007 \textit{Phys. Rev. Lett.} \textbf{98} 130602
\bibitem{SanchoPRE06}Gomez-Marin A and Sancho J M 2006 \textit{Phys. Rev.} E \textbf{74} 062102
\bibitem{Schmiedl08}Schmiedl T and Seifert U 2008 \textit{EPL} \textbf{81} 20003
\bibitem{Sekimoto97}Sekimoto K 1997 \textit{J. Phys. Soc. Jpn.} \textbf{66} 1234
\bibitem{Seifert02}Seifert U 2005 \textit{Phys. Rev. Lett.} \textbf{95} 040602
\bibitem{Schmiedl07}Schmiedl T and Seifert U 2007 \textit{Phys. Rev. Lett.} \textbf{98} 108301
\bibitem{Feynmanl}Feynman R P, Leighton R B and Sands M 1966 \textit{The Feynman Lectures on Physics} vol 1 (Reading, MA:
Addison-Wesley)
\bibitem{Prost97}J\"{u}licher F, Ajdari A and Prost J 1997 \textit{Rev. Mod. Phys.} \textbf{69} 1269
\bibitem{Astumian}Astumian R D and Bier M 1994 \textit{Phys. Rev. Lett.} \textbf{72} 1766
\bibitem{Magnasco}Magnasco M O 1993 \textit{Phys. Rev. Lett.} \textbf{71} 1477
\bibitem{Reimann96}Reimann P, Bartussek R, H\"{a}ussler R and H\"{a}nggi P 1996 \textit{Phys. Lett.} A \textbf{215} 26
\bibitem{BaoJD}Bao J D 1999 \textit{Physica} A \textbf{273} 286
\bibitem{Tu2004}Tu Z C and Ou-Yang Z C 2004 \textit{J. Phys.: Condens. Matter} \textbf{16} 1287
\bibitem{Tu2005}Tu Z C and Hu X 2005 \textit{Phys. Rev.} B \textbf{72} 033404
\bibitem{vdbrk08}van den Broeck M and van den Broeck C 2008 \textit{Phys. Rev. Lett.} \textbf{100} 130601
\bibitem{Jarzynski99}Jarzynski C and Mazonka O 1999 \textit{Phys. Rev.} E \textbf{59} 6448
\bibitem{Velasco01}Velasco S, Roco J M M, Medina A and Hern\'{a}ndez A C 2001 \textit{J. Phys. D: Appl. Phys.} \textbf{34} 1000
\bibitem{AiBQ}Ai B Q et al 2005 \textit{Eur. Phys. J.} B \textbf{48} 101
\bibitem{Zhangy}Zhang Y, Lin B H and Chen J C 2006 \textit{Eur. Phys. J.} B \textbf{53} 481
\bibitem{Parrondo}Parrondo J M R and Espa\~{n}ol P 1996 \textit{Am. J. Phys.} \textbf{64} 1125
\bibitem{Astumian99}Derenyi I and Astumian R D 1999 \textit{Phys. Rev.} E \textbf{59} R6219
\bibitem{Allahverdyan07}Allahverdyan A E, Johal R S and Mahler G 2008 \textit{Phys. Rev.} E \textbf{77} 041118

\end{thebibliography}
\end{document}